\documentclass[secnumarabic,amssymb, nobibnotes, aps, pre]{revtex4-1}

\fontsize{18pt}{\baselineskip}\selectfont

\usepackage{graphicx}% Include figure files
\usepackage{dcolumn}% Align table columns on decimal point
\usepackage{bm}% bold math
\usepackage{hyperref}% add hypertext capabilities
\usepackage{CJK}

\begin{document}

\preprint{APS/123-QED}

\title{ Modulational instability of magneto-elastic metamaterials}% Force line breaks with \\

\author{Yi S. Ding}
\email{dyi@pku.edu.cn}
 %\altaffiliation[Also at ]{School of Physics , Peking University.}%Lines break automatically or can be forced with \\
\author{Ruo-Peng Wang}%
 \email{rpwang@pku.edu.cn}
\affiliation{%
 State Key Laboratory for Mesoscopic Physics, Department of Physics, Peking University,
Beijing 100871, People's Republic of China
}%


%\affiliation{
%\affiliation{
%\affiliation{%


\date{\today}% It is always \today, today,

\begin{abstract}
We study the modulational instability of recently designed magneto-elastic metamaterials composed of elastic-mediated split-ring resonators (SRR). An effective circuit model is developed. Then four cases are studied: a dimer of SRR, one-dimension (1D) dimerized array, 1D uniform array and two-dimension layered array. It is found that except for the dimer case, all the other many-body cases can present modulational instability which may disturb the bistability (if any) under uniform assumption.

\begin{description}
\item[PACS numbers]
\verb| 42.70.Mp, 05.45.-a|
\end{description}
\end{abstract}

\maketitle

\section{introduction}
Nonlinear metamaterials (NMM) are rapidly developing fields extending the concepts of metamaterials as well as nonlinear optics.

NMM are interesting because of their extraordinary properties compared with natural nonlinear materials, including, among others, enhanced nonlinearity due to resonance \cite{pendry1999,dingjap2011}, novel phase-matching condition due to negative refractive index \cite{agranovich2004,rose2011}. Generally, NMM are inherently nonlinear lattices driven by external light radiation with subwavelength lattice constants. Under weak radiation, the responses of the lattice should be approximately described by the linear optics with appropriate modifications coming from nonlinearity. In this regime, low-order nonlinear coefficients in Taylor expansions as for natural nonlinear materials would be an appropriate description. Many recent papers work in this weak radiation regime \cite{lapine2003,agranovich2004,wen2006,joseph2010,roppo2010,zharova2005}. When radiation becomes stronger but not strong enough to damage the building materials of meta-atoms, in some cases \cite{shadrivov2006,dingjap2011} there may exist a scale of radiation intensity in/above which the lattice has such complex responses due to modulational instability that it can hardly be described by uniform optical constants (e.g., linear and/or nonlinear refractive index). That scale is characterized by the possible existence of bistability under uniform assumption regardless of the modulational instability for the case in \cite{shadrivov2006,dingjap2011,zharov2003}. Full-wave numerical simulations \cite{dingcomment} of the structures in a bistability proposal \cite{chen2011} and the results for a similar nonlinear system \cite{noskov2012} support that the bistability based on uniform assumption may be disturbed by modulational instability. In this paper we will study modulational instability for a different kind of NMM, magneto-elastic metamaterials (MEM).

Recently a new kind of metamaterials called MEM are proposed \cite{lapine2012} presenting nonlinearity arising from mechanic interactions among individual meta-atoms. In the sections below, we will first establish an effective circuit dynamic model for MEM and then investigate the properties, especially modulational instability, of four cases: (i) a dimer of split-ring resonators (SRR); (ii) a one-dimension (1D) dimerized array or an array of dimers; (iii) a 1D uniform array of SRR or a 1D polymer; (iv) layers two-dimension (2D) SRR array. We find that except for the first case all the other three many-body cases can present modulational instability above an radiation intensity threshold, which can inevitably disturb the bistability (if any) under uniform assumption.

\section{Theoretical model and an SRR dimer}\label{dimer}
A simple model based on effective circuit is developed in this section for an SRR dimer and generalized for the other many-body cases in the following sections. Linear stability analysis indicates that modulational instability can not occur for an SRR dimer.

As proposed in \cite{lapine2012}, an MEM is composed of SRR which can deviate from their original positions due to the magnetic dipole forces from their neighbors but subject to respective mechanic restoring forces. The consequence of the changing of positions is that the mutual inductance among neighbor SRR can thus be affected, rendering a feedback mechanism of the electromagnetic response of MEM, another way of saying nonlinearity.
For the simplest case, two interacting SRR sharing the same central axis, we model the SRR as effective LCR circuits and further assume that the mechanic resonance due to the restoring force and the magnetic resonance of the LCR circuit are decoupled, i.e., the frequency of the latter is much larger than that of the former, and also that the damping of the mechanic oscillation is very fast compared to the varying rate of envelope of magnetic response. In other words, the balanced position of each SRR can simply be determined by the averaged magnetic forces (related to the envelope of magnetic response) it experiences and the restoring force coefficients. Possible electric interactions are not taken into account in this paper for simplicity. These assumptions are consistent with the ones made in \cite{lapine2012} and are intended here for deriving simple dynamic equations for the system.

In Appendix \ref{A}, we obtain the dynamic equations for a dimer under slowly-varying approximation
\begin{equation}
2i\Omega\frac{d\tilde{Q}_{1,2}}{d\tau}+(-\Omega^2+i\gamma\Omega+1)\tilde{Q}_{1,2}-\kappa\Omega^2
\left[1+\Omega^2\mbox{Re}(\tilde{Q}_1\tilde{Q}_2^*)\right]\tilde{Q}_{2,1}=u_{1,2},\label{8A}
\end{equation}
where $\tilde{Q}$, $\Omega$, $\tau$, $\gamma$, $\kappa$ and $u$ are dimensionless quantities related to charge, frequency, time, damping, mutual inductance and electromotive force, respectively. Included in Appendix \ref{A} are more details on the approximation to derived Eq.(\ref{8A}) and the definitions of the dimensionless quantities.

The stationary states of Eq.(\ref{8}) then satisfy
\begin{equation}
(-\Omega^2+i\gamma\Omega+1)\tilde{Q}_{1,2}-\kappa\Omega^2
\left[1+\Omega^2\mbox{Re}(\tilde{Q}_1\tilde{Q}_2^*)\right]\tilde{Q}_{2,1}=u_{1,2}.\label{9}
\end{equation}

Under uniform radiation $u_1=u_2=u$, the uniform stationary states $\tilde{Q}_1=\tilde{Q}_2=\tilde{Q}$ satisfy
\begin{equation}
(-\Omega^2+i\gamma\Omega+1)\tilde{Q}-\kappa\Omega^2
\left[1+\Omega^2|\tilde{Q}|^2\right]\tilde{Q}=u.\label{10}
\end{equation}

\begin{figure}[h]\centering
\includegraphics[height=6cm,width=8cm]{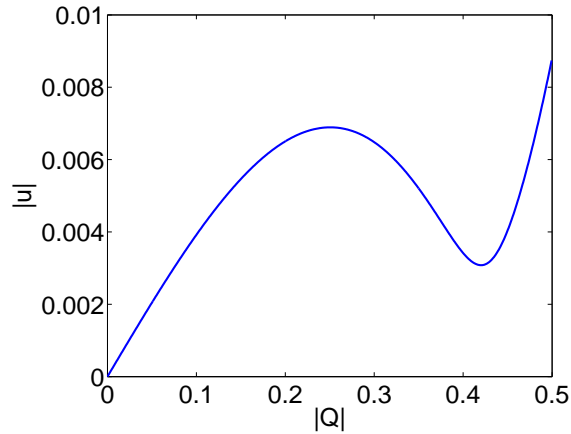}
\caption{A typical bistability curve for the uniform response of a dimer. Parameters: $\Omega=0.73$, $\gamma=0.01$, $\kappa=0.8$}\label{fig1}
\end{figure}

Bistability can easily be obtained from the above uniform equation. See Fig.\ref{fig1} for an example.

\subsection{Modulational instability for an SRR dimer}
Will the two SRR always respond uniformly under uniform radiation $u_1=u_2$? In order to answer this question we should investigate the modulational instability of the uniform stationary states in Eq.(\ref{10}). Before we do linear stability analysis for the uniform equation (\ref{10}), we want to point out that the stationary equations (\ref{9}) without making uniform assumption do not have nonuniform solutions \footnote{One can just subtract one with the other equation in Eq.(\ref{9}) to verify this observation.}, implying no modulational instability could occur for a dimer. Linear stability analysis in Appendix \ref{B} also confirms this speculation.

\section{1D Dimerized SRR array}\label{1DD}
Since a dimer can only have uniform response, it will be convenient to treat it as a unit nonlinear element rather than two SRR. A question is: can modulational instability occur when an array of dimers are arranged along a line sharing the same central axis (or plane), i.e., for a 1D dimerized SRR array which models the experimental structures in Fig.5 of \cite{lapine2012}. In this section, we investigate the modulational instability of such an array.

If each dimer in such an array is treated as a unit nonlinear element and the interactions between neighbors are considered to be linear, the dynamic equation can be easily written according to Eq.(\ref{8A}),
\begin{equation}
2i(1+\kappa)\Omega\frac{dQ_n}{d\tau}+(-\Omega^2+i\gamma\Omega+1-\kappa\Omega^2-\kappa\Omega^4|Q_n|^2)Q_n-\kappa_d\Omega^2(Q_{n-1}+Q_{n+1})=u.\label{11}
\end{equation}
where $\kappa_d$ is the linear interaction coefficient between nearest dimers. We have denoted $\tilde{Q}$ by $Q$ for simplicity and employed the nearest-neighbor approximation.

Under uniform assumption and periodic boundary conditions, the stationary uniform states can be obtained from
\begin{equation}
(-\Omega^2+i\gamma\Omega+1-\kappa\Omega^2-\kappa\Omega^4|Q|^2)Q-2\kappa_d\Omega^2Q=u.\label{12}
\end{equation}

\begin{figure}[h]\centering
\includegraphics[height=6cm,width=8cm]{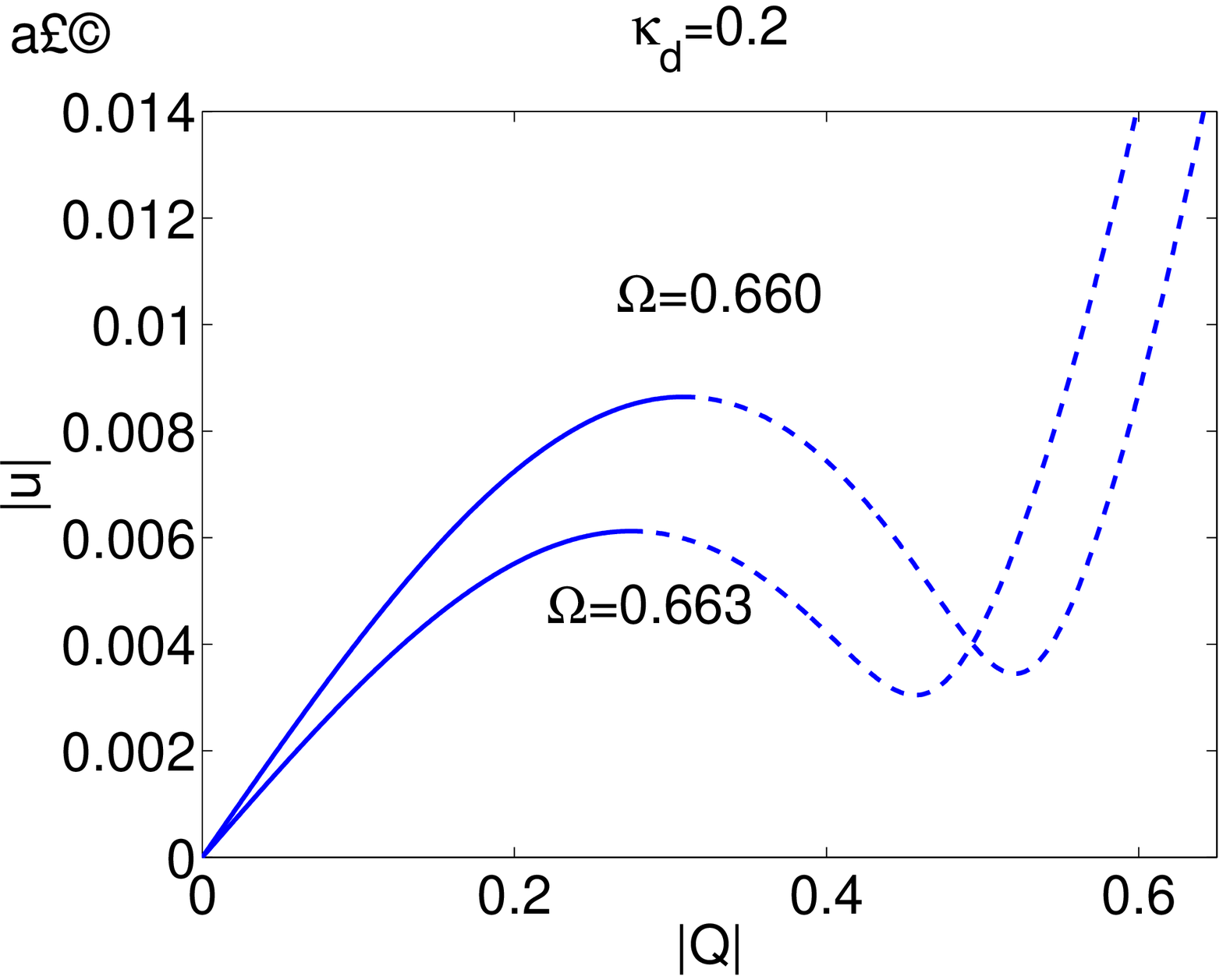}
\includegraphics[height=6cm,width=8cm]{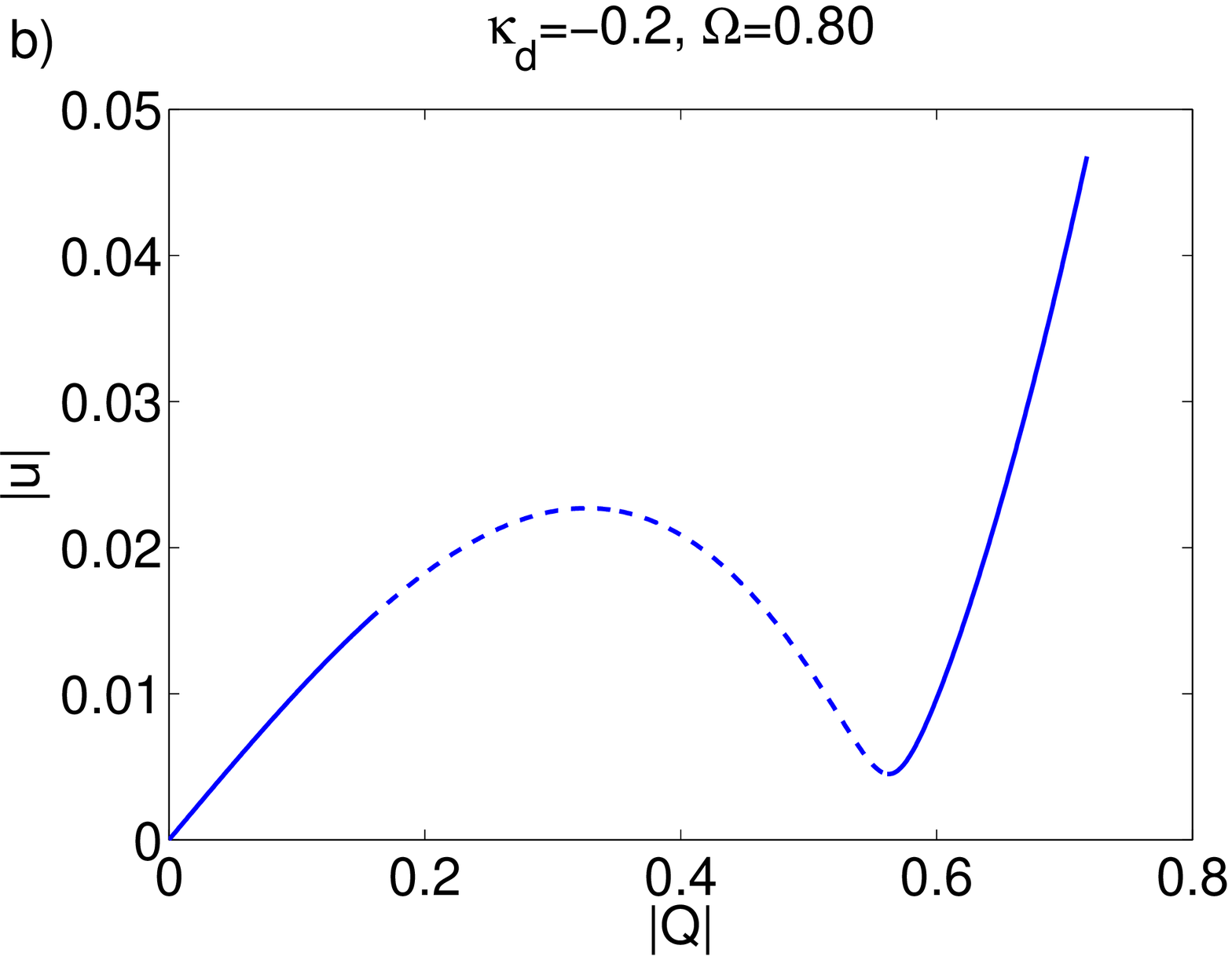}
\caption{Typical uniform responses of a dimerized array. Parameters:  $\gamma=0.01$, $\kappa=0.8$. The dashed portions of the curves are instable ranges.}\label{fig2}
\end{figure}

Bistability can also present for Eq.(\ref{12}) as shown in Fig.\ref{fig2}.

\subsection{Modulational instability of a dimerized array}
We do stability analysis in Appendix \ref{C} for this case. The growth rate for each Fourier component of the fluctuation is calculated. Unlike the dimer, a dimerized array can indeed present modulational instability which may disturb the upper or lower branch of the bistability depending on the sign of the interaction coefficient $\kappa_d$ (A minus $\kappa_d$ means that the dimers share the same central plane). The unstable ranges are denoted by the dashed line in Fig.\ref{fig2}.

\begin{figure}[h]\centering
\includegraphics[height=6cm,width=8cm]{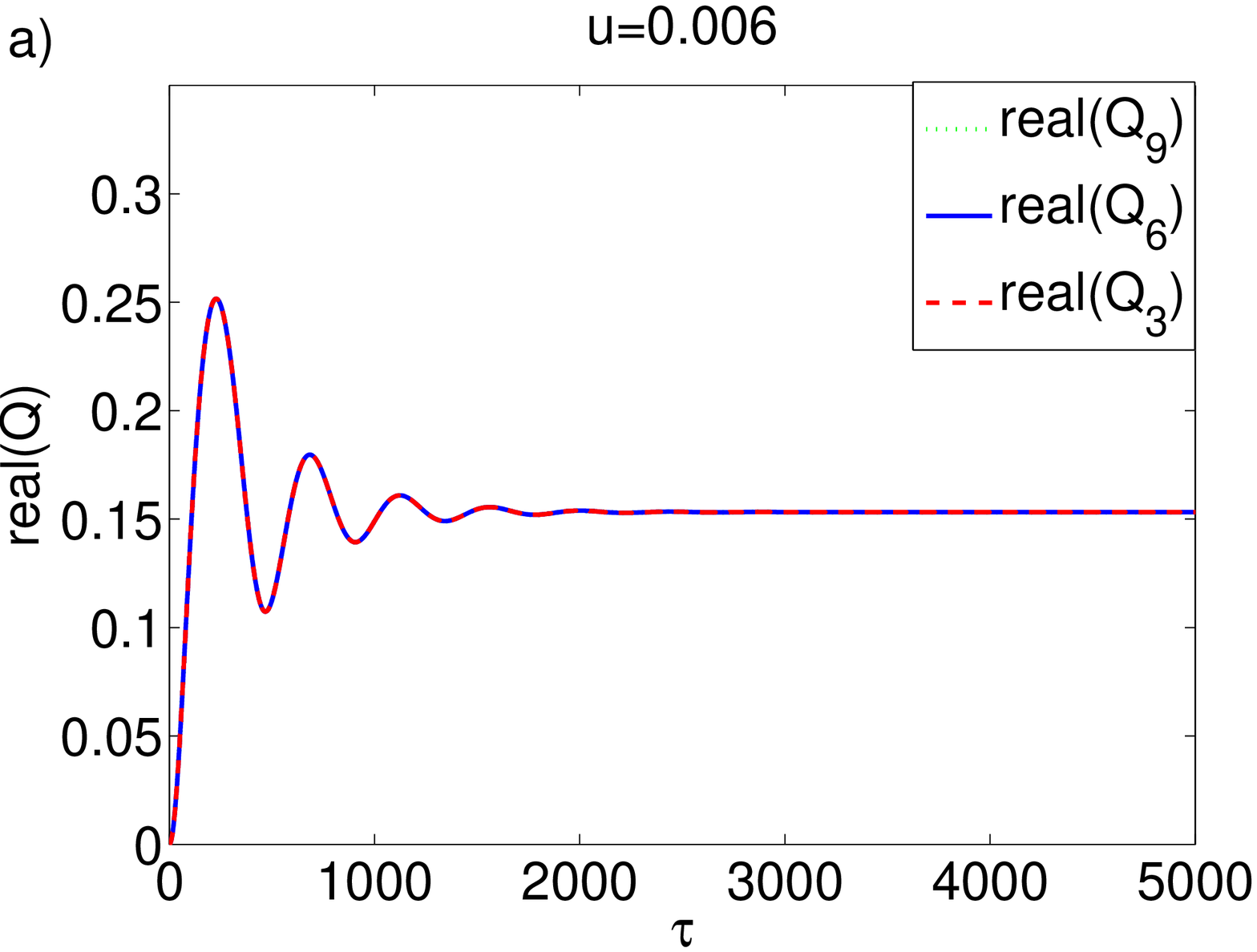}
\includegraphics[height=6cm,width=8cm]{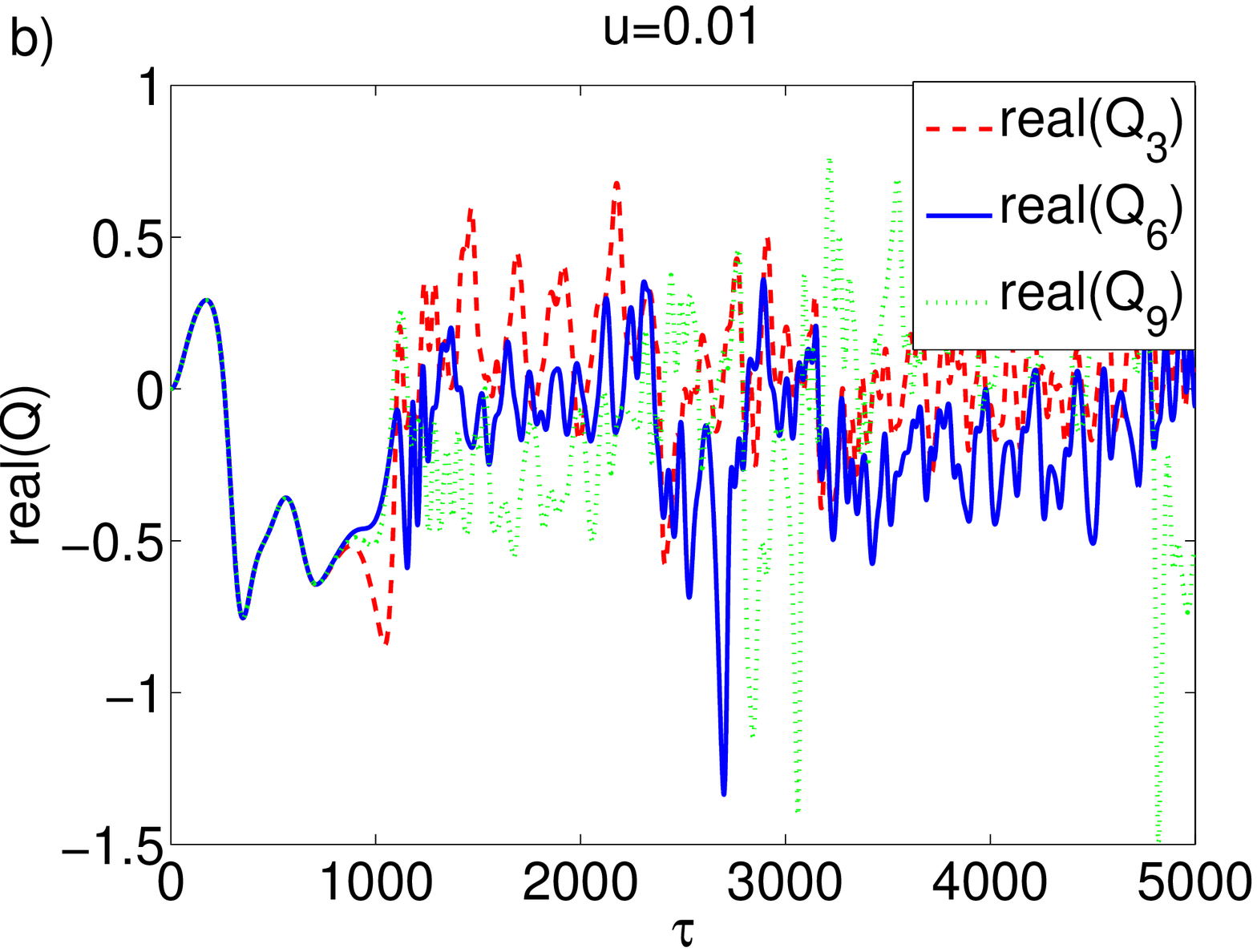}
\caption{Time-domain responses of three dimers in the dimer array. Parameters: $\kappa=0.8$, $\gamma=0.01$, $\kappa_d=0.2$, $\Omega=0.660$, $u=0.006$ (a), $u=0.01$ (b).}\label{fig3}
\end{figure}

\begin{figure}[h]\centering
\includegraphics[height=4cm,width=5cm]{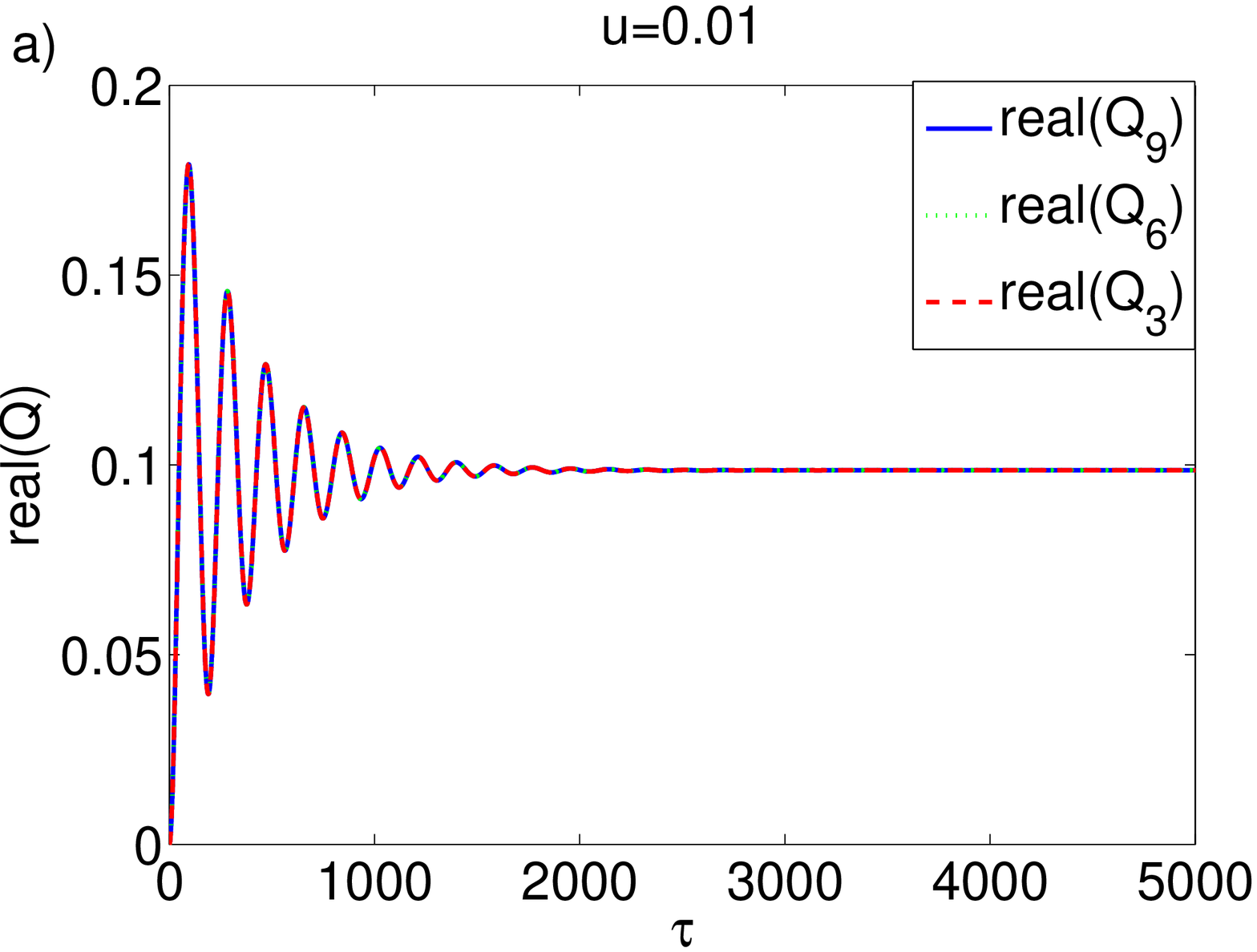}
\includegraphics[height=4cm,width=5cm]{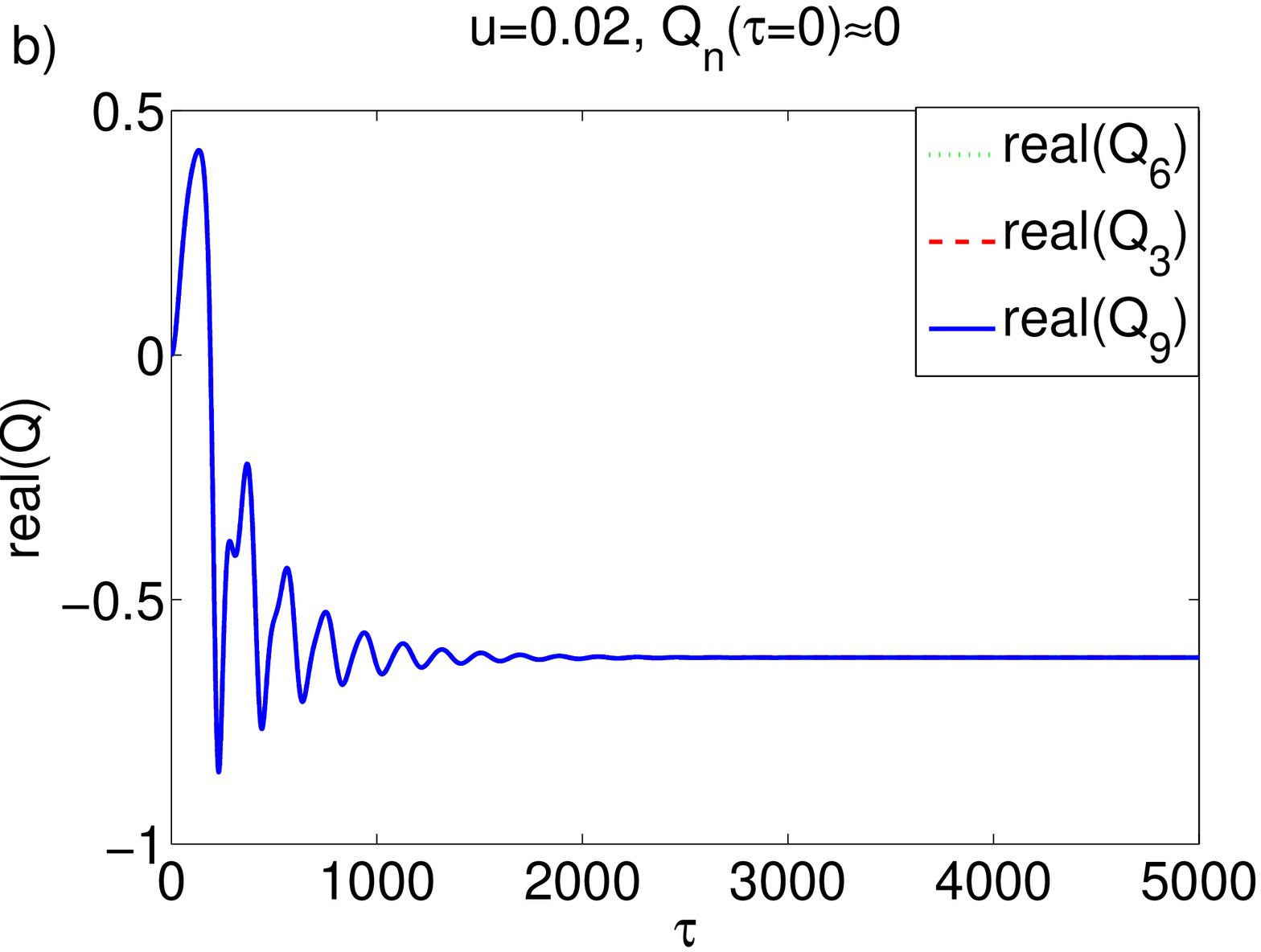}
\includegraphics[height=4cm,width=5cm]{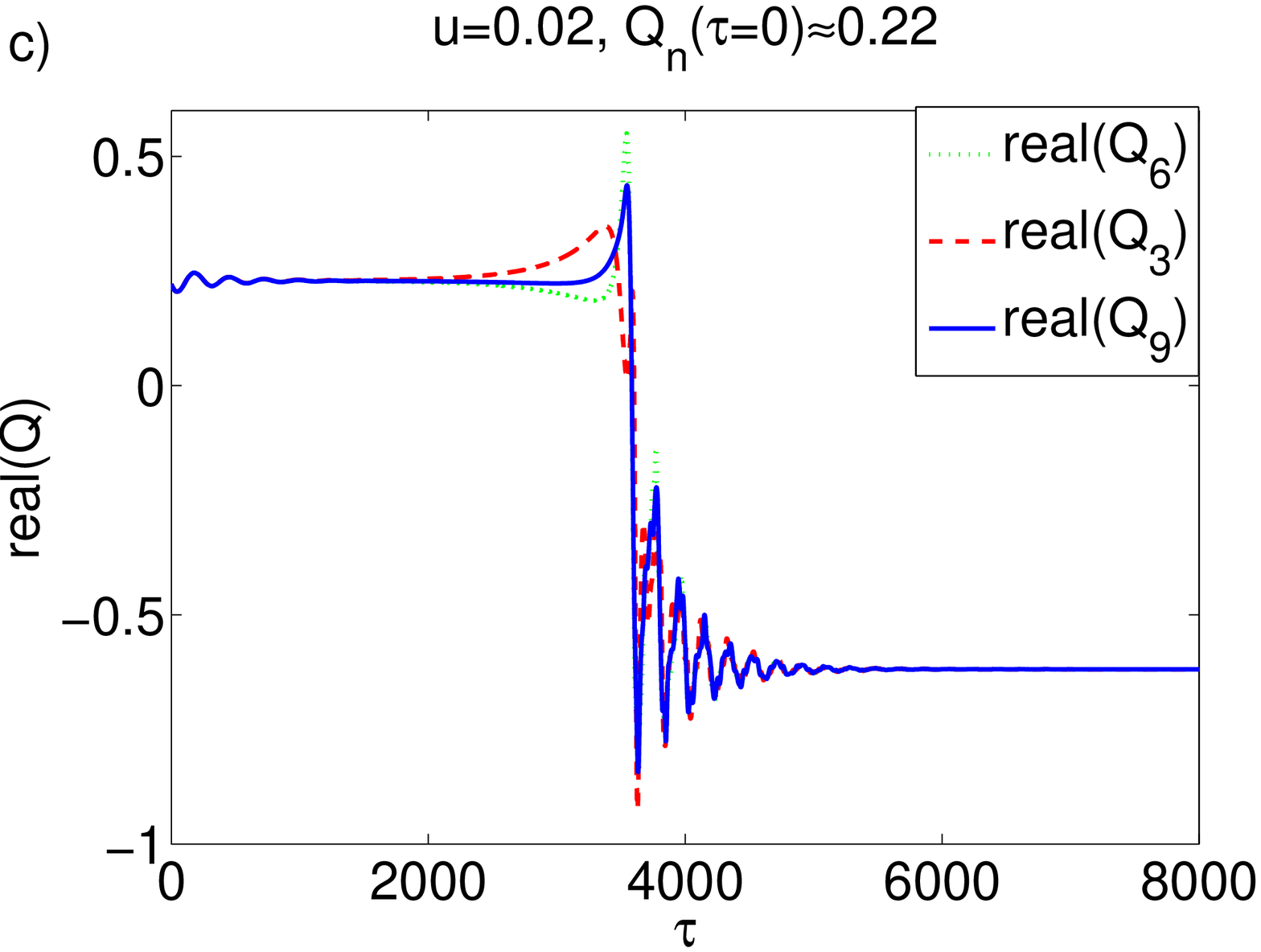}
\caption{Time-domain responses of three dimers in the dimer array. Parameters: $\kappa=0.8$, $\gamma=0.01$, $\kappa_d=-0.2$, $\Omega=0.80$, $u=0.01$ (a), $u=0.02$ (b and c). The initial conditions are $Q_n(\tau=0)\approx0$ (a and b) and $Q_n(\tau=0)\approx0.22$ (c)}\label{fig4}
\end{figure}

In order to see to what extent modulational instability can disturb the uniform response, we do time-domain simulations by directly integrating Eq.(\ref{11}). In order to trigger the modulational instability, we introduce tiny fluctuations to the initial states of the array, i.e., $Q_n(\tau=0)$ randomly distribute within $[0,\,\,\,\,10^{-6}]$. The radiation is turned on at $\tau=0$ and is stationary thereafter. We consider an array containing nine dimers and monitor the time-domain responses of the third, sixth and ninth one in Fig.\ref{fig3} and Fig.\ref{fig4}.

In Fig.\ref{fig3}, the simulations are for the $\kappa_d=0.2$, $\Omega=0.660$ case in Fig.\ref{fig2}. The upper branch of the bistability is completely covered by the unstable range. We test two radiation intensities corresponding to $u=0.006$ (corresponding to a stable uniform state outside the unstable range) and $u=0.01$ (inside the unstable range). For $u=0.01$, the modulational instability manifests itself as chaos.

In Fig.\ref{fig4}, the simulations are for the $\kappa_d=-0.2$, $\Omega=0.80$ case in Fig.\ref{fig2}. Part of the lower branch of the bistability is unstable as shown in Fig.\ref{fig2}. We test two radiation intensities corresponding to $u=0.01$ and $u=0.02$. For the latter, we test two initial conditions, $Q_n(\tau=0)\approx0$ and $Q_n(\tau=0)\approx0.22$. For $u=0.01$, no modulational instability exists as we expected from Fig.\ref{fig2}. However, when $u=0.02$ and $Q_n(\tau=0)\approx0$, the system without experiencing the possible nonuniform responses directly skip to the upper branch which does not suffer from modulational instability. If we prepare the initial states inside the unstable range $Q_n(\tau=0)\approx0.22$ with tiny fluctuations, we can indeed observe nonuniform responses due to modulational instability for a while, but the system will finally leak into the upper uniform branch anyway.

\section{1D uniform array}\label{1DU}
In the 1D array in the previous section, the nonlinear interactions occur pairwise. In this section, we consider the case in which the nonlinear interaction is considered to be uniform, i.e., each SRR has equal nonlinear interactions with its neighbors. More precisely, the balanced position of each SRR is determined by the magnetic forces of all its neighbors (rather than only that of its partner as in the dimerized array case) and the positions of SRR in turn affect the mutual inductances. In Appendix \ref{D}, we derive the dynamic equations for this case under the slowly-varying and nearest-neighbor approximations,
\begin{eqnarray}
&&2i\Omega\frac{d\tilde{Q_n}}{d\tau}+(-\Omega^2+i\Omega\gamma+1)\tilde{Q}_n=u_n+\kappa\Omega^2[1+\Omega^2\mbox{Re}
(\tilde{Q}_n\tilde{Q}_{n+1}^*-2\tilde{Q}_{n-1}\tilde{Q}_{n}^*+\tilde{Q}_{n-1}\tilde{Q}_{n-2}^*)]\tilde{Q}_{n-1}\nonumber\\
&&+\kappa\Omega^2[1+\Omega^2\mbox{Re}(\tilde{Q}_{n+1}\tilde{Q}_{n+2}^*-2\tilde{Q}_{n}\tilde{Q}_{n+1}^*+\tilde{Q}_{n}\tilde{Q}_{n-1}^*)]\tilde{Q}_{n+1}.\label{13}
\end{eqnarray}

The stationary states under uniform assumption and periodic boundary conditions are
\begin{eqnarray}
&&[-(1+2\kappa)\Omega^2+i\Omega\gamma+1]\tilde{Q}=u.\label{14}
\end{eqnarray}

We observe that the relation between the response $Q$ and the radiation $u$ is linear and thus no bistability can exist. That is because the array is infinitely long (due to the periodic boundary conditions) so that the net magnetic force vanishes for each SRR. But it is not the case for an array of finite length, e.g., a dimer in Sec.\ref{dimer}, or for a nonuniform array, e.g., a dimerized array in Sec.\ref{1DD}.

\begin{figure}[h]\centering
\includegraphics[height=6cm,width=8cm]{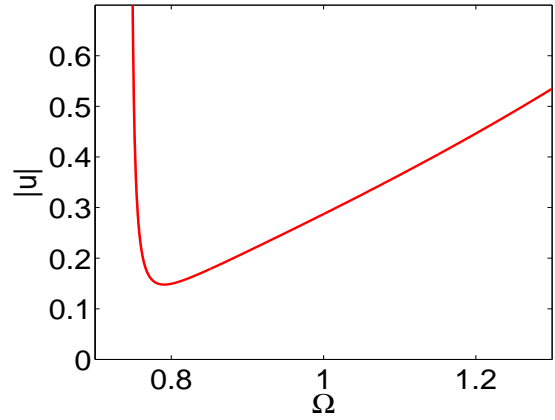}
\caption{Boundary between stable and unstable range for 1D uniform array.}\label{fig5}
\end{figure}

While the uniform responses do not present nonlinear effects, we can still study possible modulational instability of the uniform linear responses. That is analytically done in Appendix \ref{D} in which we have obtained the growth rate for each Fourier component of the fluctuation. Only when all the modes of fluctuations decay, the uniform response is stable. In Fig.\ref{fig5}, we find the boundary between the stable and unstable range for $\kappa=0.5$ and $\gamma=0.1$. We can see that there exists a critical frequency (about $\Omega=0.74$) below which no modulational instability occurs.

\begin{figure}[h]\centering
\includegraphics[height=4cm,width=5cm]{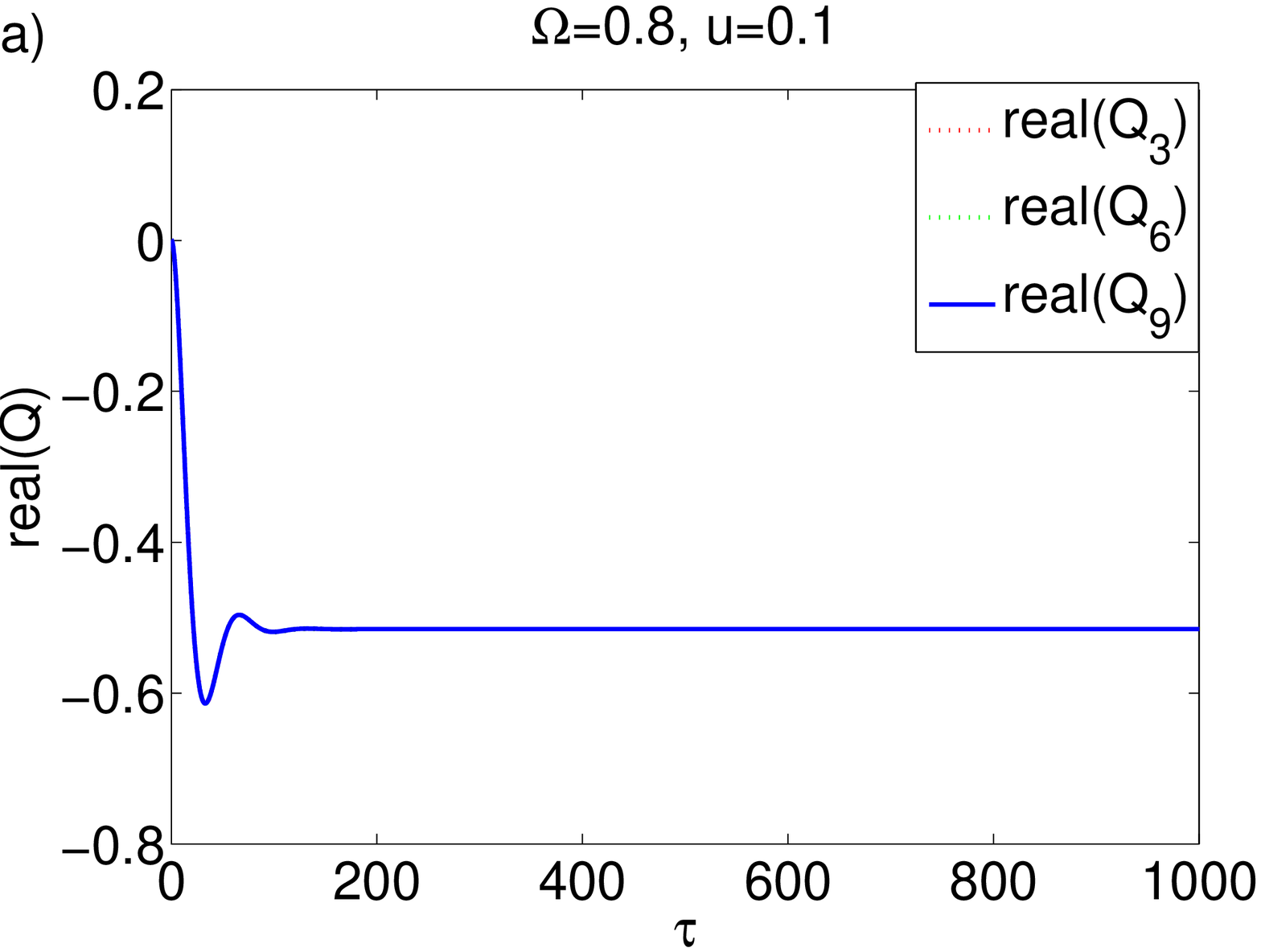}
\includegraphics[height=4cm,width=5cm]{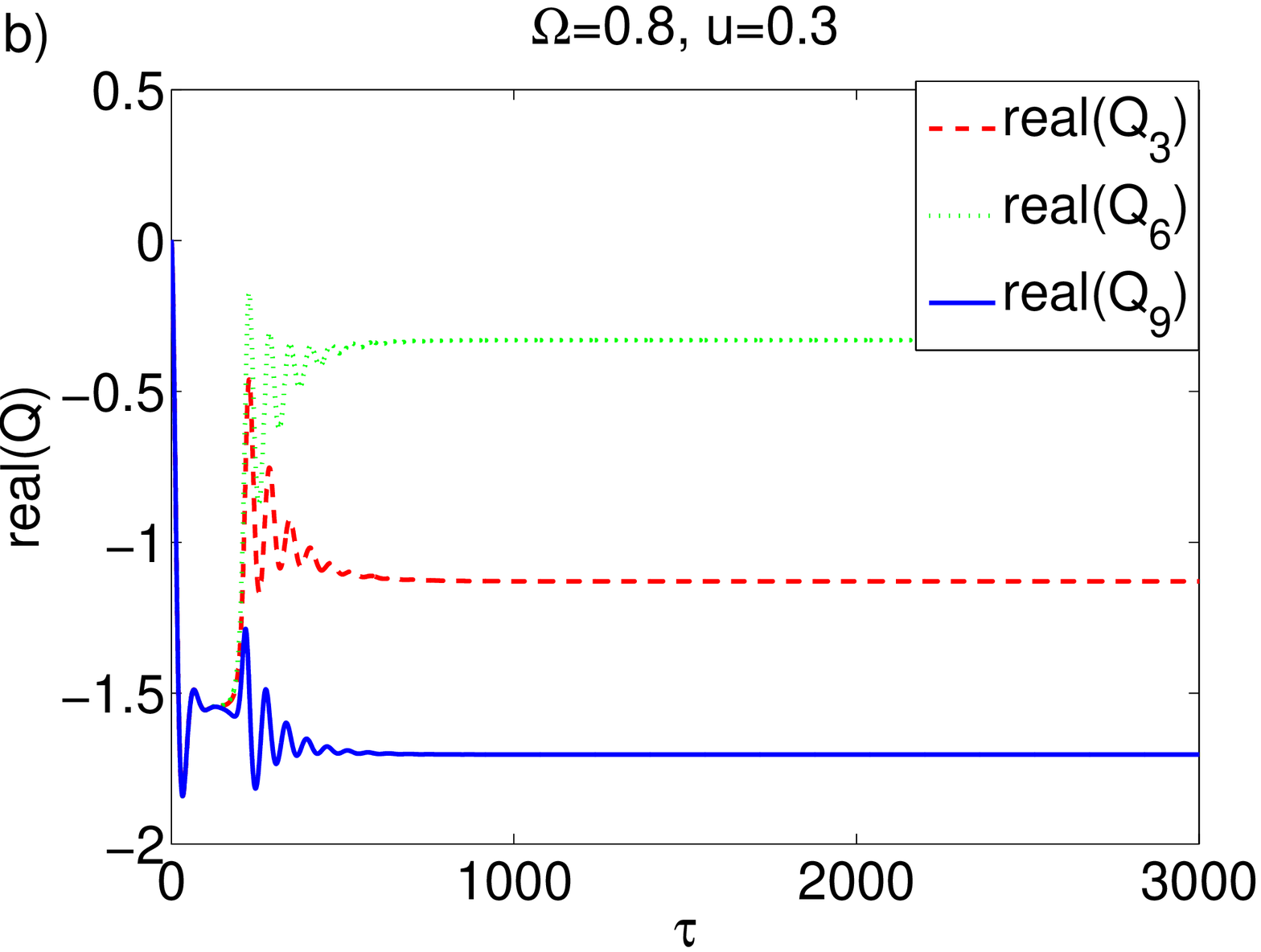}
\includegraphics[height=4cm,width=5cm]{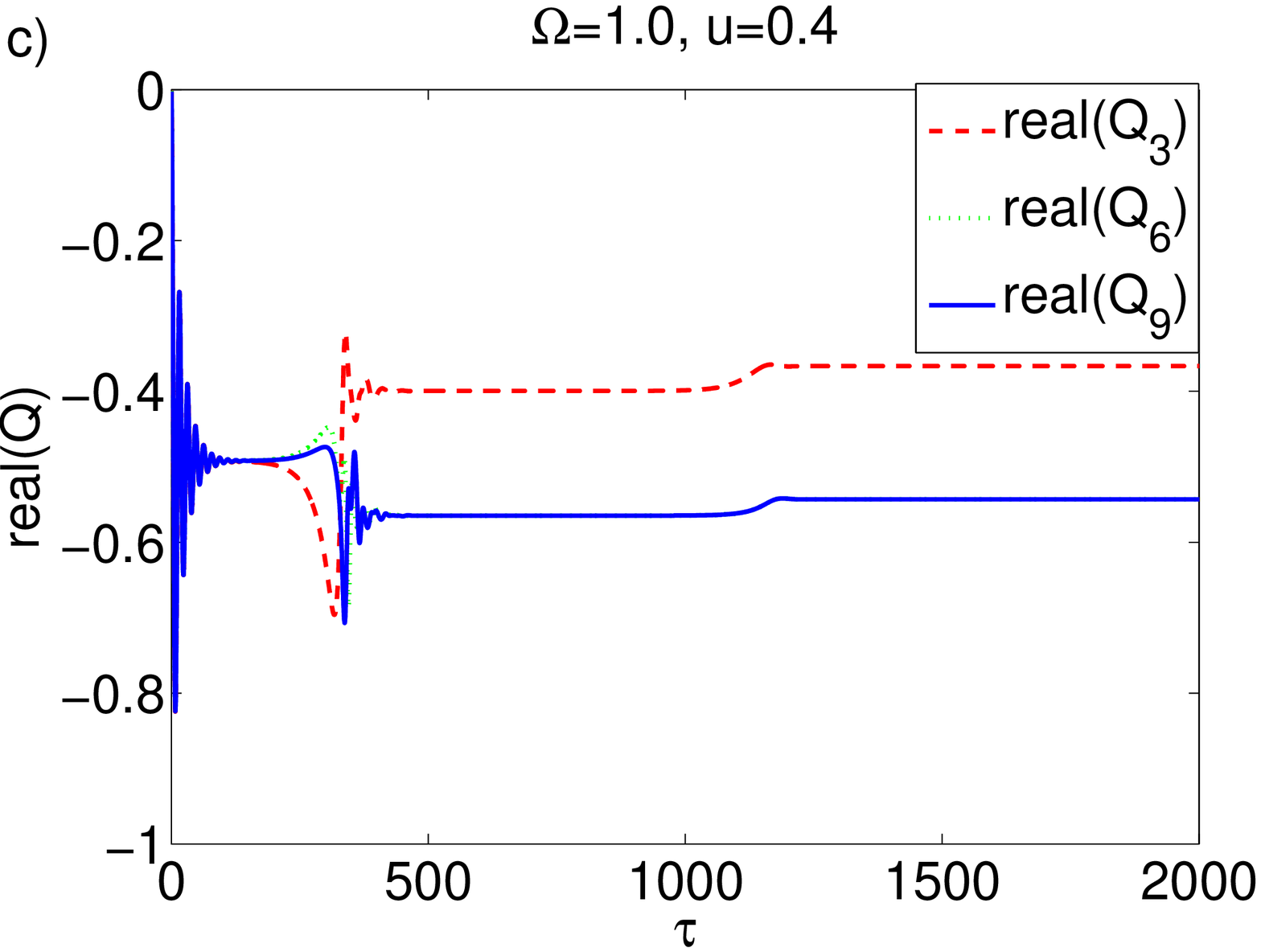}
\caption{Time-domain responses of three SRR of a uniform array containing ten SRR with periodic boundary conditions. Parameters: $\kappa=0.4$, $\gamma=0.1$.}\label{fig6}
\end{figure}

Next, we do time-domain simulations to show how modulational instability manifest itself in this case. In Fig.\ref{fig6}, we show three cases, one in the stable range, the other two in the unstable range. We find that the system can have nonuniform but stationary states in the unstable range for those specific cases.

\section{2D layered SRR array}\label{2D}
We then consider the 2D structures originally proposed in \cite{lapine2012} (Fig.1). We construct a similar model for this 2D layered structures in Appendix \ref{E} as the cases in previous sections. The main simplifications are nearest-neighbor approximations, slowly-varying approximations, and Taylor approximations for mutual inductances and magnetic forces. The the main deviation of these approximations from the model in \cite{lapine2012} is that the former gives simple bistability of uniform responses as shown below while the the latter gives a more complicated one. But our model can capture the discrete nature of that structure which is essential for studying modulational instability. The calculations below show that modulational instability may also occur in the uniform bistability range just as in the dimerized-array case in Sec.\ref{1DD}.

\begin{figure}[h]\centering
\includegraphics[height=6cm,width=8cm]{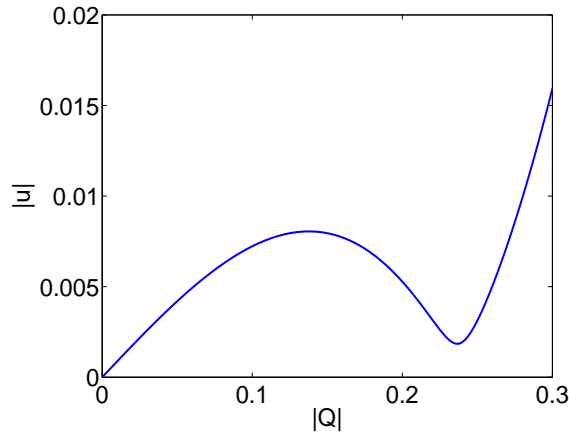}
\caption{Bistability under uniform assumption for 2D layered magneto-elastic metamaterials. Parameters: $g=0.5$, $m_0=-0.1$, $m_{10}=0.5$, $m_{20}$=0.1, $m_{21}=0.1$, $\omega=0.78$, $\gamma=0.01$. The detailed definitions of the dimensionless parameters can be found in Appendix \ref{E}. }\label{fig7}
\end{figure}

\begin{figure}[h]\centering
\includegraphics[height=6cm,width=8cm]{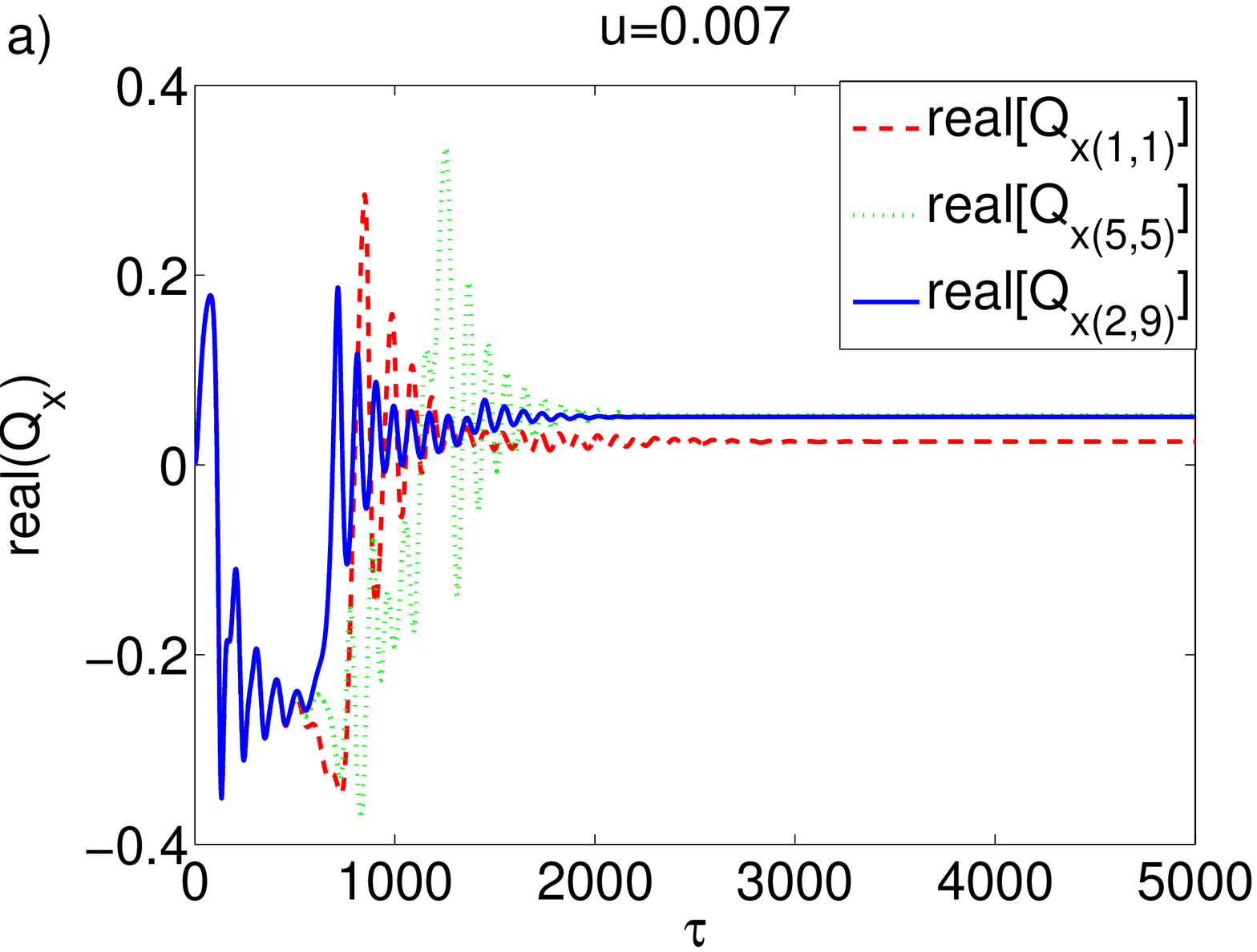}
\includegraphics[height=6cm,width=8cm]{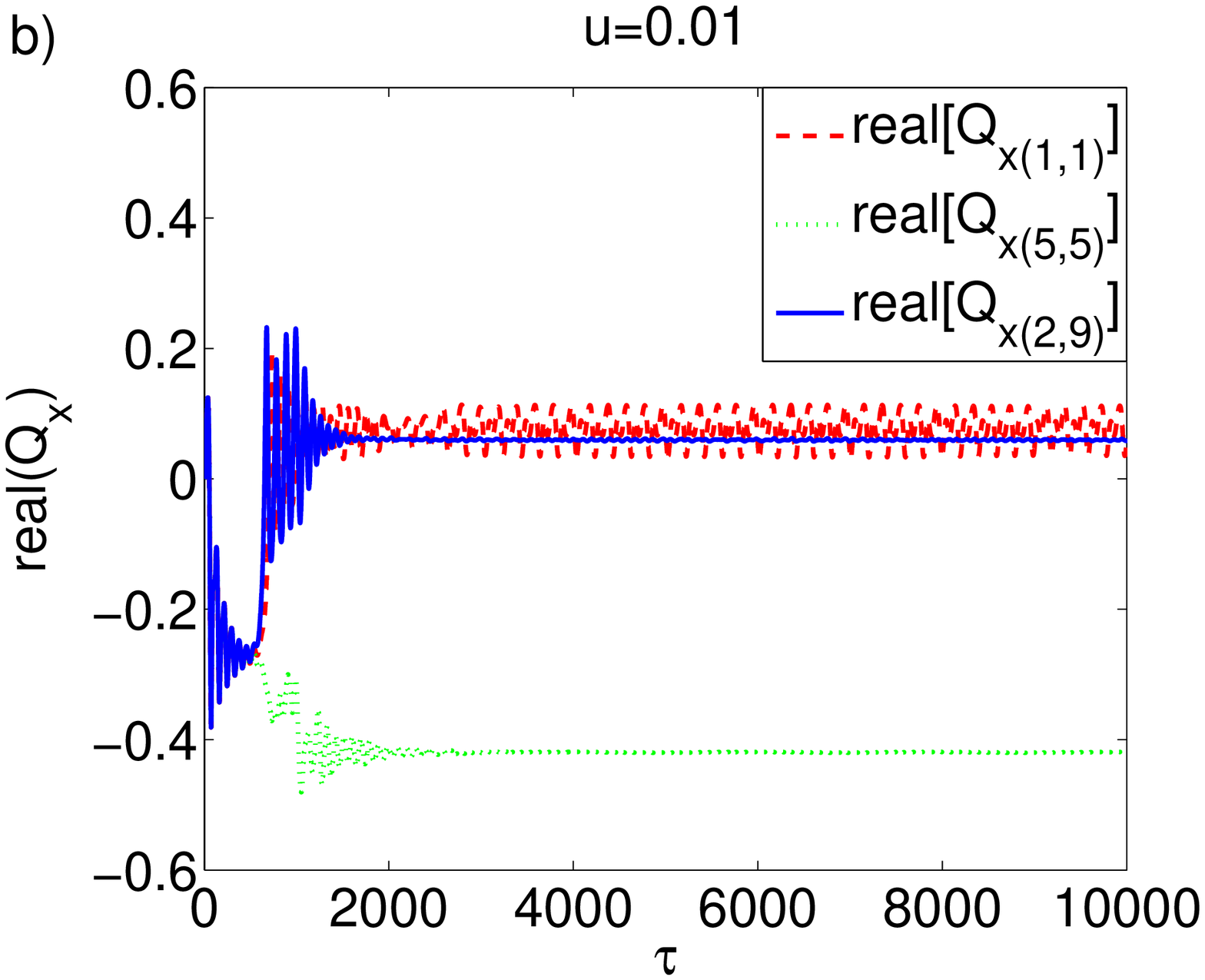}
\includegraphics[height=6cm,width=8cm]{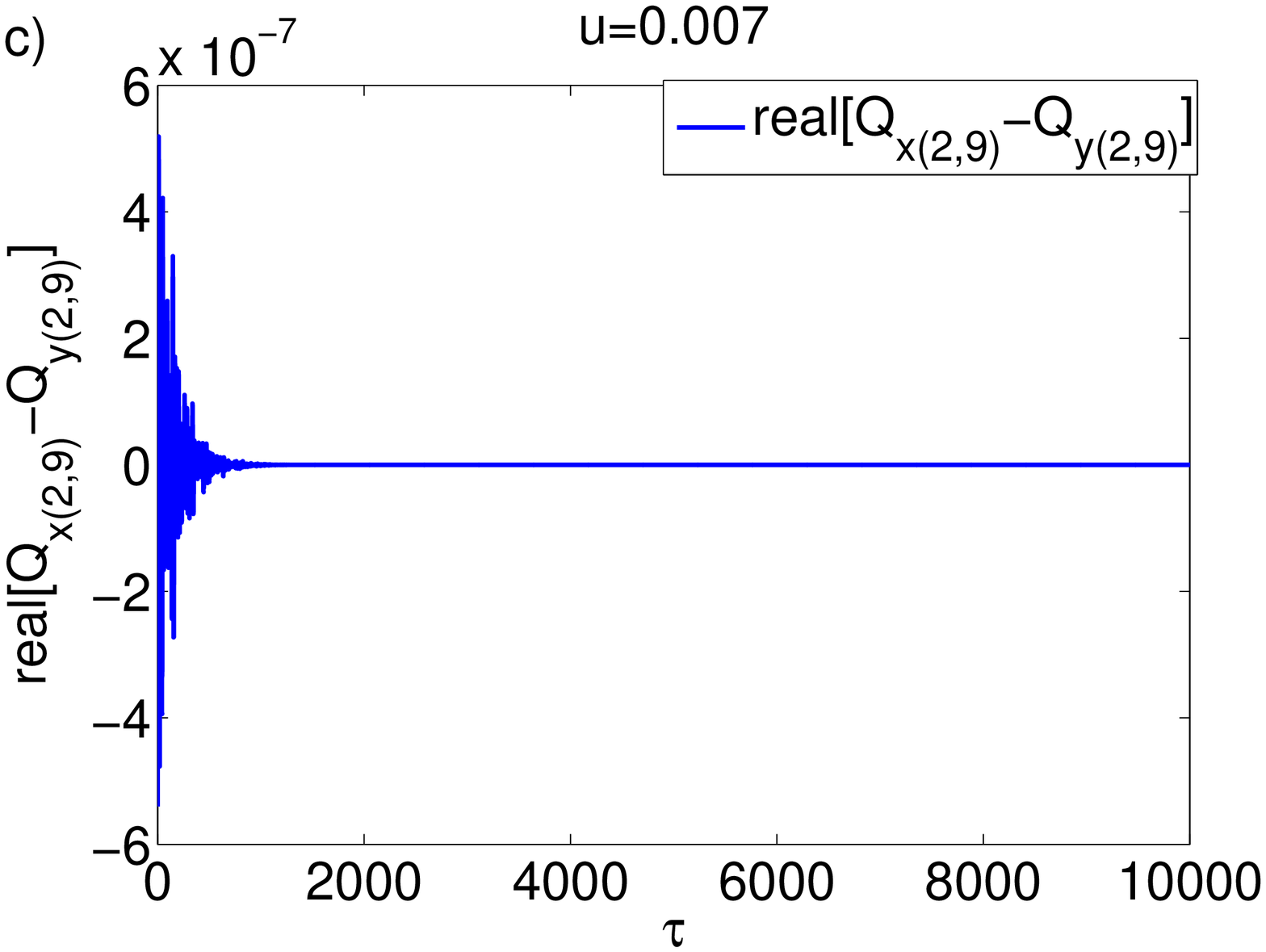}
\caption{Time-domain responses of three SRR of a 2D layered array containing $2\times10\times10$ SRR with periodic boundary conditions in the directions parallel to the two layers. The parameters are the same as in Fig.\ref{fig7}. Two intensities with modulational instability are tested, and limit-cycle responses are observed. The figure (c) shows the difference between two counterparts at the site (2,9).}\label{fig8}
\end{figure}

Based on the simplifications mentioned above, we derive dynamic equations for this 2D nonlinear structure in Appendix \ref{E}. Bistable responses under uniform assumptions can occur in this model (see Fig.\ref{fig7}). But when doing time-domain integrations in Fig.\ref{fig8}, we find that modulational instability manifest itself as nonuniform (e.g., limit-cycle) responses disturbing the uniform bistability. We also find that despite the complex responses within one of the two layers, the responses of counterparts in the two layers keep close to each other in the tested case, similar to the dimer in Sec.\ref{dimer}.

\section{Discussion}
We discuss why modulational instability seems to usually accompany the uniform bistability for nonlinear many-body systems in this paper. Uniform bistability is characterized by a middle unstable branch, which according to the Fourier analysis in Appendix \ref{B} is the unstable range for the uniform (or $k=0$) fluctuations. Since for a infinitely large system the wave vector of fluctuations continuously ranges from $k=0$ to $k=\pi/a$ ($a$ is the lattice constant), it is natural that the whole unstable range, the union of unstable ranges for all wave vectors, can be obtained by continuously extending the original unstable middle branch. It can also be seen from the Fourier analysis in Appendix \ref{B} that the range of bistability disturbing by modulational instability is proportional to the coupling coefficients between neighbor elements. Therefore, for negligible coupling, modulational instability should not be an important factor.

It is worth pointing out that a recent publication \cite{noskov2012} which studies the modulational instability accompanying possible uniform bistability also verify the arguments above in a completely different system.

\section{Conclusion}
We develop an effective circuit model for a recently designed magneto-elastic metamaterials and study the modulational instability of them. Four cases are studied: a dimer, 1D dimerized array, 1D uniform array, and 2D layered array. We find that a dimer do not present modulational instability and can be treated as a unit nonlinear element. All the other three many-body cases can have modulational instability showing chaotic, stationary but nonuniform, or limit-cycle responses. The modulational instability seems to usually accompany the uniform bistability for the 1D uniform array and 2D layered array.

\appendix
\section{Dynamic equations for an SRR dimer}\label{A}
The dynamic equations for the two SRR based on effective circuit model are
\begin{equation}
L\frac{dq_{1,2}^2}{dt^2}+R\frac{dq_{1,2}}{dt}+\frac{q_{1,2}}{C}=\mathcal{E} e^{i\omega t}-M\frac{dq_{2,1}^2}{dt^2}\label{1}
\end{equation}
where $q_{1,2}$ is the respective charge hold by the capacitors of the two SRR, $L$ is the self-inductance, $R$ is the resistance, $C$ is the capacitance, $M$ is the mutual inductance, and $\mathcal{E}e^{i\omega t}$ is the effective electromotive force.

Since the mutual inductance is determined by the averaged magnetic force between SRR, we are only interested in the amplitudes $Q_{1,2}$ of the charges $q_{1,2}=Q_{1,2}(t)e^{i\omega t}$. Then we should derive the dynamic equations for $Q_{1,2}$ under slowly-varying approximation as in \cite{shadrivov2006}, which finally read
\begin{equation}
2i\omega L \frac{dQ_{1,2}}{dt}+(-\omega^2L+i\omega R+\frac{1}{C})Q_{1,2}-M\omega^2Q_{2,1}=\mathcal{E}_{1,2}\label{2}
\end{equation}

Next, we remember that the mutual inductance $M$ is a function of the magnetic response. First, we approximate $M$ by its Taylor expansion reserving only the first two orders,
\begin{equation}
M\approx M_0+\frac{\partial M}{\partial b}(b-b_0),\label{3}
\end{equation}
where $b$ is the distance between the two SRR, $b_0$ is the distance for zero magnetic force, and $M_0$ is the mutual inductance for $b=b_0$.

According to Hooke's law,
\begin{equation}
K(b-b_0)=-<F(b,I_1,I_2)>\label{4}
\end{equation}
where $K$ is the coefficient of the restoring force, $F$ is the instantaneous magnetic force as a function of the two currents $I_{1,2}$ , and $<\cdots>$ denotes the time-averaged value.

The instantaneous magnetic force should be of the form,
\begin{equation}
F(b,I_1,I_2)=I_1I_2f(b)\approx I_1I_2 f(b_0),\label{5}
\end{equation}
while the time-averaged value should thus be
\begin{equation}
<F(b,I_1,I_2)>\approx \frac{1}{2}\mbox{Re}(I_1I_2^*)f_0\label{6}
\end{equation}
where $I_{1,2}$ are complex amplitudes of the currents and $f_0=f(b_0)$. Note that we have made further simplification that $f(b)$ is approximated by $f(b_0)$.

Then after combining Eq.(\ref{3}-\ref{6}) the mutual inductance $M$ can be expressed as a function of the complex currents
\begin{equation}
M=M_0-\frac{\partial M}{\partial b_0}\frac{f_0\mbox{Re}(I_1I_2^*)}{2K}.\label{7}
\end{equation}

At this point we are ready to write down the full dimensionless dynamic equations for the SRR dimer under the slowly-varying approximation and other simplifications mentioned above,
\begin{equation}
2i\Omega\frac{d\tilde{Q}_{1,2}}{d\tau}+(-\Omega^2+i\gamma\Omega+1)\tilde{Q}_{1,2}-\kappa\Omega^2
\left[1+\Omega^2\mbox{Re}(\tilde{Q}_1\tilde{Q}_2^*)\right]\tilde{Q}_{2,1}=u_{1,2},\label{8}
\end{equation}
where we have defined $|I_c|^2=\left|\frac{2K M_0}{f_0\frac{\partial M}{\partial b_0}}\right|$, $\omega_0=\frac{1}{\sqrt{LC}}$, $\tilde{Q}_{1,2}=\frac{\omega_0Q_{1,2}}{|I_c|}$, $\tau=\omega_0t$, $\Omega=\omega/\omega_0$, $\gamma=\omega_0RC$, $\kappa=\frac{M_0}{L}$ and $u_{1,2}=C\mathcal{E}_{1,2}\omega_0/|I_c|$.

\section{Linear stability analysis for an SRR dimer}\label{B}
We do linear expansion of the dynamic equations (\ref{8A}) near the uniform stationary states obtained from Eq.(\ref{10}). Substituting $\tilde{Q}_{1,2}=\tilde{Q}+\delta_{1,2}$ to Eq.(\ref{8A}) where $\tilde{Q}$ is a uniform stationary state, we arrive at the linear dynamic equations for fluctuations $\delta_{1,2}$
\begin{eqnarray}
&&2i\Omega\frac{d\delta_{1,2}}{d\tau}+(-\Omega^2+i\gamma\Omega+1)\delta_{1,2}-\kappa\Omega^2(1+\Omega^2|\tilde{Q}|^2)\delta_{2,1}\nonumber\\
&&-\kappa\frac{\Omega^4}{2}(\delta_2^*\tilde{Q}^2+\delta_1|\tilde{Q}|^2+\delta_2|\tilde{Q}|^2+\delta_1^*\tilde{Q}^2)=0.\label{B1}
\end{eqnarray}
Subtract one of the two equations above with the other,
\begin{eqnarray}
&&2i\Omega\frac{d(\delta_1-\delta_2)}{d\tau}+(-\Omega^2+i\gamma\Omega+1)(\delta_{1}-\delta_2)
-\kappa\Omega^2(1+\Omega^2|\tilde{Q}|^2)(\delta_{2}-\delta_1)=0.\label{B2}
\end{eqnarray}
Clearly, this equation only has decaying solution for $\delta_1-\delta_2$, which means that the uniform states from Eq.(10) are stable under nonuniform fluctuations.

\section{Linear stability analysis for a dimerized array}\label{C}
From the dynamic equations (\ref{11}), the dynamics of fluctuations around uniform stationary states $Q_n(t)=Q_s+\delta_n(t)$ should read.
\begin{equation}
2i(1+\kappa)\Omega\frac{d\delta_n}{d\tau}+(C_1+2C_2|Q_s|^2)\delta_n+C_2Q_s^2\delta_n^*+C_3(\delta_{n-1}+\delta_{n+1})=0
\end{equation}
where $C_1=-\Omega^2+i\gamma\Omega+1-\kappa\Omega^2$, $C_2=-\kappa\Omega^4$, $C_3=-\kappa_d\Omega^2$. Or

\begin{equation}
2i(1+\kappa)\Omega\frac{d\delta_n}{d\tau}+A\delta_n+B\delta_n^*+C(\delta_{n-1}+\delta_{n+1})=0
\end{equation}

The above equation is linear and has periodic symmetry, so the eigenmodes should be uniform plane waves. Let $\delta_n=\alpha e^{ikn}+\beta e^{-ikn}$. Then
\begin{eqnarray}
\frac{d\alpha}{d\tau}&=&-i\eta[(A+2C\cos k)\alpha+B\beta^*]\\
\frac{d\beta^*}{d\tau}&=&-i\eta[-B^*\alpha-(A^*+2C\cos k)\beta^*]
\end{eqnarray}
where $1/i\eta=2i(1+\kappa)\Omega$. Note that $\eta<0$.

The eigen values of the coefficient matrix in the right-hand side are
\begin{eqnarray}
\lambda&=&\eta\left[\gamma\Omega\pm\sqrt{|B|^2-|A+2C\cos k|^2+\gamma^2\Omega^2}\right]\\
&=&\eta\left[\gamma\Omega\pm\sqrt{(\kappa\Omega^4|Q_s|^2)^2-(-\Omega^2+1-\kappa\Omega^2-2\kappa\Omega^4|Q_{s}|^2-2\kappa_d\Omega^2\cos k)^2}\right]
\end{eqnarray}

 The $\lambda>0$ indicates unstable states , which lie in the range
\begin{equation}
\kappa\Omega^4|Q_s|^2\in[\frac{1}{3}\left(2a-\sqrt{a^2-3b^2}\right),\,\,\,\,\frac{1}{3}\left(2a+\sqrt{a^2-3b^2}\right)]
\end{equation}
where $a=-\Omega^2+1-\kappa\Omega^2-2\kappa_d\Omega^2\cos k$, $b=\gamma\Omega$.

\section{1D uniform array and modulational instability}\label{D}
We consider a periodic chain of SRRs with nonlinear magneto-elastic interactions and periodic boundary conditions.

The deviations of each SRR from the original balanced positions are denoted by $x_{i}$, $i=0,1,\cdots$.
According to Hooke's law, and the nearest-neighbor approximation,
\begin{equation}
Kx_n=\frac{1}{2}\mbox{Re}(I_nI_{n+1}^*-I_{n-1}I_n^*)f_0.
\end{equation}
We have further approximated the magnetic force by its zeroth order.

Mutual inductance between the n'th and n+1'th is
\begin{equation}
M(x_n,x_{n+1})=M_0+\frac{\partial M}{\partial x}(x_{n+1}-x_n)
\end{equation}

Then the stationary equation is
\begin{eqnarray}
&&-\omega^2LQ_n+i\omega RQ_n+\frac{Q_n}{C}=\mathcal{E}+\left[M_0+\frac{\partial M}{\partial x}\frac{f_0}{2K}\mbox{Re}(I_nI_{n+1}^*-2I_{n-1}I_{n}^*+I_{n-1}I_{n-2}^*)\right]\omega^2Q_{n-1}\nonumber\\
&&+\left[M_0+\frac{\partial M}{\partial x}\frac{f_0}{2K}\mbox{Re}(I_{n+1}I_{n+2}^*-2I_{n}I_{n+1}^*+I_{n}I_{n-1}^*)\right]\omega^2Q_{n+1},
\end{eqnarray}
where we have assumed nearest-neighbor approximation for the mutual inductance.

The above equation can be recast into a dimensionless form in a similar way to the dimer case in Appendix \ref{A},

\begin{eqnarray}
&&(-\Omega^2+i\Omega\gamma+1)\tilde{Q}_n=u_n+\kappa\Omega^2[1+\mbox{Re}(i_ni_{n+1}^*-2i_{n-1}i_{n}^*+i_{n-1}i_{n-2}^*)]\tilde{Q}_{n-1}\nonumber\\
&&+\kappa\Omega^2[1+\mbox{Re}(i_{n+1}i_{n+2}^*-2i_{n}i_{n+1}^*+i_{n}i_{n-1}^*)]\tilde{Q}_{n+1}
\end{eqnarray}
or
\begin{eqnarray}
&&(-\Omega^2+i\Omega\gamma+1)\tilde{Q}_n=u_n+\kappa\Omega^2[1+\Omega^2\mbox{Re}
(\tilde{Q}_n\tilde{Q}_{n+1}^*-2\tilde{Q}_{n-1}\tilde{Q}_{n}^*+\tilde{Q}_{n-1}\tilde{Q}_{n-2}^*)]\tilde{Q}_{n-1}\nonumber\\
&&+\kappa\Omega^2[1+\Omega^2\mbox{Re}(\tilde{Q}_{n+1}\tilde{Q}_{n+2}^*-2\tilde{Q}_{n}\tilde{Q}_{n+1}^*+\tilde{Q}_{n}\tilde{Q}_{n-1}^*)]\tilde{Q}_{n+1}\label{D1}
\end{eqnarray}
since $i_n=i\Omega\tilde{Q}$.

The dynamic equations can be easily obtained from Eq.(\ref{D1}) under slowly-varying approximation,
\begin{eqnarray}
&&2i\Omega\frac{d\tilde{Q_n}}{d\tau}+(-\Omega^2+i\Omega\gamma+1)\tilde{Q}_n=u_n+\kappa\Omega^2[1+\Omega^2\mbox{Re}
(\tilde{Q}_n\tilde{Q}_{n+1}^*-2\tilde{Q}_{n-1}\tilde{Q}_{n}^*+\tilde{Q}_{n-1}\tilde{Q}_{n-2}^*)]\tilde{Q}_{n-1}\nonumber\\
&&+\kappa\Omega^2[1+\Omega^2\mbox{Re}(\tilde{Q}_{n+1}\tilde{Q}_{n+2}^*-2\tilde{Q}_{n}\tilde{Q}_{n+1}^*+\tilde{Q}_{n}\tilde{Q}_{n-1}^*)]\tilde{Q}_{n+1}\label{U_dynamic}
\end{eqnarray}

In order to study the modulational instability, dynamics for fluctuations, $Q=Q_s+\delta_n$ is also needed.
\begin{eqnarray}
&&2i\Omega\frac{d\delta_n}{d\tau}=-(-\Omega^2+i\Omega\gamma+1)\delta_n+\kappa\Omega^2(\delta_{n-1}+\delta_{n+1})+\kappa\Omega^2([\clubsuit]+[\spadesuit])Q
\end{eqnarray}
where
\begin{eqnarray}
[\clubsuit]=\Omega^2\mbox{Re}[Q^*(\delta_{n-2}-\delta_{n-1}-\delta_{n}+\delta_{n+1})]
\end{eqnarray}
and
\begin{eqnarray}
[\spadesuit]=\Omega^2\mbox{Re}[Q^*(\delta_{n-1}-\delta_{n}-\delta_{n+1}+\delta_{n+2})]
\end{eqnarray}

Then
\begin{eqnarray}
&&2i\Omega\frac{d\delta_n}{d\tau}=-(-\Omega^2+i\Omega\gamma+1)\delta_n+\kappa\Omega^2(\delta_{n-1}+\delta_{n+1})+\\
&&\kappa\Omega^4\mbox{Re}[Q^*(\delta_{n-2}-2\delta_{n}+\delta_{n+2})]Q
\end{eqnarray}

Let $\delta_n=\alpha e^{ikn}+\beta e^{-ikn}$. We have
\begin{eqnarray}
&&2i\Omega\dot\alpha =-(-\Omega^2+i\gamma\Omega+1)\alpha+2\kappa\Omega^2\alpha\cos k +\kappa\Omega^4[Q^*(\cos 2k-1)\alpha+Q(\cos2k-1)\beta^*]Q\\
&&2i\Omega\dot\beta=-(-\Omega^2+i\gamma\Omega+1)\beta+2\kappa\Omega^2\beta\cos k +\kappa\Omega^4[Q^*(\cos 2k-1)\beta+Q(\cos2k-1)\alpha^*]Q
\end{eqnarray}

We take the complex conjugation of the second equation
\begin{eqnarray}
&&2i\Omega\dot\alpha =-(-\Omega^2+i\gamma\Omega+1)\alpha+2\kappa\Omega^2\alpha\cos k +\kappa\Omega^4[Q^*(\cos 2k-1)\alpha+Q(\cos2k-1)\beta^*]Q\\
&&-2i\Omega\dot\beta^* =-(-\Omega^2-i\gamma\Omega+1)\beta^*+2\kappa\Omega^2\beta^*\cos k +\kappa\Omega^4[Q(\cos 2k-1)\beta^*+Q^*(\cos2k-1)\alpha]Q^*
\end{eqnarray}

Reform
\begin{eqnarray}
&&\dot\alpha=\frac{1}{i\eta}(A\alpha+B\beta^*)\\
&&\dot\beta=\frac{1}{i\eta}(-B^*\alpha-A^*\beta^*)
\end{eqnarray}
where $i\eta'=2i\Omega$ ($\eta'>0$), $A=-(-\Omega^2+i\gamma\Omega+1)+2\kappa\Omega^2\cos k+\kappa\Omega^4|Q|^2(\cos2k-1)$ and $B=\kappa\Omega^4Q^2(\cos2k-1)$.

The characteristic polynomial and its solutions
\begin{equation}
(A-\lambda)(-A^*-\lambda)+|B|^2=0
\end{equation}
\begin{equation}
\lambda'=\frac{\lambda}{i\eta'}=\frac{1}{\eta'}\left(-\gamma\Omega\pm\sqrt{\gamma^2\Omega^2+|B|^2-|A|^2}\right)
\end{equation}

\begin{equation}
\lambda'=\frac{1}{\eta'}\left(-\gamma\Omega\pm\sqrt{-(\Omega^2-1+2\kappa\Omega^2\cos k)^2-2(\Omega^2-1+2\kappa\Omega^2\cos k)\kappa\Omega^4|Q|^2(\cos 2k-1)}\right)
\end{equation}

(We may use the two symbols $Q$ and $\tilde{Q}$ interchangeably in this paper, but it is clear that in a dimensionless equation, $Q$ means $\tilde{Q}$.)

\section{Dynamic equations for a 2D layered magneto-elastic metamaterial}\label{E}

We consider a two-dimension magneto-elastic metamaterial as in Fig.1 of \cite{lapine2012}. It contains two layers (x, y) of split-ring resonators. We will make the following approximations. The mutual magnetic inductances are restricted to its counterpart in the other layer, four nearest neighbors in the same layer and the neighbors of the counterpart. The mutual mechanic interactions are restricted to its counterpart and its neighbors' counterparts. The mechanic interaction coefficients (e.g., the $f$ function in Appendix \ref{A}) do not depend on the distance between the split-ring resonators.

The displacement of the SRR at the site (m,n) of the x layer is denoted by $x_{m,n}$. Then
\begin{eqnarray}
Kx_{m,n}=\frac{1}{2}\mbox{Re}[(I_{y(m,n)}f_0+I_{y(m+1,n)}g_0+I_{y(m-1,n)}g_0+I_{y(m,n+1)}g_0+I_{y(m,n-1)}g_0)I_{x(m,n)}^*],
\end{eqnarray}
with a similar relation for the y layer.

Mutual inductances:

\begin{equation}
M(x_{m,n},x_{m,n-1})=M(x_{m,n},x_{m,n+1})=M(x_{m,n},x_{m-1,n})=M(x_{m,n},x_{m+1,n})=M_{0}<0
\end{equation}

\begin{equation}
M(x_{m,n},y_{m,n-1})=M_{20}+M_{21}(x_{m,n}-y_{m,n-1})
\end{equation}

\begin{equation}
M(x_{m,n},y_{m,n+1})=M_{20}+M_{21}(x_{m,n}-y_{m,n+1})
\end{equation}

\begin{equation}
M(x_{m,n},y_{m-1,n})=M_{20}+M_{21}(x_{m,n}-y_{m-1,n})
\end{equation}

\begin{equation}
M(x_{m,n},y_{m+1,n})=M_{20}+M_{21}(x_{m,n}-y_{m+1,n})
\end{equation}

\begin{equation}
M(x_{m,n},y_{m,n})=M_{10}+M_{11}(x_{m,n}-y_{m,n})
\end{equation}

The static equation is
\begin{equation}
-\omega^2LQ_{x(m,n)}+i\omega RQ_{x (m,n)}+\frac{Q_{x (m,n)}}{C}=\mathcal{E}+\omega^2\sum_\alpha M_\alpha Q_\alpha.
\end{equation}
where the summation $\sum_\alpha$ runs over all the neighbors and counterparts of $Q_{x(m,n)}$.

The dynamic equation under slowly-varying approximation is
\begin{equation}
i\omega\dot{Q}_{x(m,n)}-\omega^2LQ_{x( m,n)}+i\omega RQ_{x(m,n)}+\frac{Q_{x(m,n)}}{C}=\mathcal{E}+\omega^2\sum_\alpha M_\alpha Q_\alpha,
\end{equation}
with a corresponding equation for the y layer.

The nondimensionalization procedure is similar to that in Appendix \ref{A} with some additional definitions: setting $f_0=1$, $M_{1,1}=1$, and defining $g=g_0/f_0$, $m_0=M_0/L$, $m_{10}=M_{10}/L$, $m_{20}=M_{20}/L$, $m_{21}=M_{21}/M_{11}$.

%\bibliography{reference}% Produces the bibliography via BibTeX.
%Merlin.mbs v4.21 2009-07-09.

\end{document}